\documentclass[journal=jpcafh,manuscript=article]{achemso}
\usepackage{amsmath}
\usepackage{amsmath,amssymb,amsfonts,latexsym,graphicx}
\usepackage{geometry}
\usepackage{indentfirst}
\usepackage{fancyhdr}
\usepackage{subfigure}
\usepackage{algorithm}
\usepackage{algorithmic}
\usepackage{titletoc}
\usepackage{titlesec}
\usepackage{booktabs}
\usepackage{threeparttable}
\usepackage{multirow}
\usepackage{color}
\usepackage{rotating}
\usepackage[version=3]{mhchem} 
\setkeys{acs}{maxauthors = 0} 

\title{Theoretical Study of  the Anisotropy Spectra of the Valine Zwitterion and Glyceraldehyde}

\author{Jie Su}

\author{Bingbing Suo}
\affiliation{Shaanxi Key Laboratory for Theoretical Physics Frontiers, Institute of Modern Physics, Northwest University, Xi'an, Shaanxi 710027, China}
\email{bsuo@nwu.edu.cn}

\author{Patrick Cassam-Chena\"{i}}
\affiliation{Universit\'e C\^ote d'Azur, CNRS, LJAD, France.}
\email{cassam@unice.fr}


\begin{document}

%
%
%
%
%

\begin{abstract}
The electronic absorption (EA), circular dichroism (ECD), and anisotropy spectra of the L-valine zwitterion and D-glyceraldehyde are calculated by time-dependent density functional theory (TDDFT) with the M06-2X and B3LYP functionals. It is found that the absorption and ECD spectra from TDDFT/M06-2X agree well with experimental results measured from the amorphous film of L-valine. Moreover, the calculations reproduce all three major peaks observed in the experimental anisotropy spectra. For D-glyceraldehyde, the TDDFT/M06-2X calculations indicate that the excitation wavelengths of the first excited state of 32 stable conformers distribute from 288 to 322 nm, giving rise to two ECD peaks with opposite signs centered at 288 nm and 322 nm. The very weak absorption of the first excited state (S1) induces two high peaks in the anisotropy spectra of D-glyceraldehyde, which should be seen in future experimental works.
\end{abstract}


\section{ Introduction }
The basic units of biochemical macromolecules of a given class, such as nucleic acids and proteins, are molecules with identical stereo-chemical L- or D-configurations, when chiral. For example, all chiral amino acids found in natural, ribosomally synthesized oligopeptides and proteins have a left-handed configuration, while nucleic acids possess a backbone composed exclusively of D-sugars.  It is still unclear why life on Earth demonstrates such an intriguing asymmetry at the molecular level. A promising hypothesis is that  photochemical processes in extraterrestrial space  induced synthesis and enantioenrichment of the chiral organic molecules,\cite{griesbeck2002asymmetric, meierhenrich2008book} which were delivered to the early Earth by meteorites or comets.\cite{life2019} Experiments that simulate the photochemical evolution of  interstellar ice analogs have synthesized several amino acids.\cite{bernstein2002, caro2002amino} Other studies have identified the formation of DNA and RNA related molecules such as aldehydes, ribose, and related sugars in ultra-violet irradiated  interstellar ice analogs.\cite{Meinert2015, Meinert2016ribose} Furthermore, enantiomeric excesses (\textit{ee}) of amino acids were obtained in experiments where circularly polarized (CP) light, known to exist in astrophysical environments, was introduced.\cite{Nuevo2007, AminoacidCPLee2011, Meinert2014}  All these experimental evidence support an astrochemical origin of chiral, building block molecules for life on Earth.

The enantioselection of the chiral molecules exposed to CP light arises from the different absorption capacity of the enantiomers. That is, both left-hand and right-hand enantiomers absorb CP light, however, one has a slightly smaller absorption coefficient than the other, hence it will be photo-destroyed less rapidly and an initial racemic mixture will be enantioenriched. The reaction rate of photo-induced \textit{ee} is mainly determined by the anisotropy factor $g$ (also known as the absorption dissymmetry factor),\cite{Balavoine1974}
\begin{align*}
g=2\frac{\varepsilon_R-\varepsilon_S}{\varepsilon_R+\varepsilon_S}
\end{align*} 
in which $\varepsilon_R$ and $\varepsilon_S$ are the molar absorption coefficients of the right-hand and left-hand enantiomers, respectively. 

The  anisotropy factor is a dimensionless quantity introduced by Kuhn in 1930 \cite{Kuhn1930}. Although early authors seem to have followed Kuhn's terminology \cite{Condon1937}, the same quantity is alternatively referred to as the "dissymmetry factor'' in the more recent literature \cite{berova2000circular,Anisotropic2014,Chirality2014Review,PCCP2016solvation}. One major property of the anisotropy factor is that it does not depend on the amount of sample molecules, so, neither upon the molecule concentration nor upon the thickness of the sample. Moreover, when comparing experimental and theoretical spectra, the anisotropy spectra has been found to be "a more discerning tool'' than the corresponding ECD spectra.\cite{Chirality2014Review,PCCP2016solvation} Experiments using synchrotron radiation have extended the ECD spectra deep into the vacuum ultraviolet (VUV) region,\cite{miles2003calibration, Hoffmann2007synchrotron, hussain2008design, Tanaka2009synchrotron, ECDValin2009, meierhenrich2010circular, Tanaka2010} reporting the anisotropy spectra of several amino acids.\cite{Meinert2012anisotropy} The derivation of the anisotropy spectra from experimental absorption and ECD spectra usually depends upon the segmentation of the latter into slices of somewhat arbitrary finite sizes to compute their ratio. In contrast, a theoretical anisotropy factor can be calculated straightforwardly as
\begin{equation}\label{eqn:anspec}
g=\frac{4R}{D}
\end{equation} 
where $R$ is the rotational strength and $D$ is the dipole strength of a chiral molecule, as discussed by Schellman more than forty years ago.\cite{Schellman1975}. Comparison of experimental anisotropy spectra with theoretical ones 
\cite{Meinert2014, Meinert2015oleb, Chirality2014Review, PCCP2016solvation,Anisotropic2014} has become a  complement to the conventional chiral-optical techniques\cite{dijerassi1960, UVECD1962, burke1972low, berova2000circular} for analysing the structure of chiral molecules.

It is still a challenging task to obtain a good agreement between a theoretical and an experimental anisotropy spectrum.\cite{Chirality2014Review,PCCP2016solvation} In the present study, there are two major difficulties. Firstly, experimental anisotropy spectra have been evaluated by measuring absorption and ECD spectra of amorphous films on an \ce{MgF2} window,  and nothing is known about the orientational freedom of the molecules in the film. Theoretical calculations are usually carried out for isolated molecules, sometimes for molecules in solution but we have no reason to believe that the latter modelling is more relevant to the amorphous solid state than the former, since experiments have revealed strong differences between spectra in solution and in amorphous film.\cite{Bredehoft2014gsolvent}. Moreover, different solvents may give quite different spectral features.\cite{PCCP2016solvation} Secondly, the molecules  in this study are floppy, that is to say, they have many accessible, stable geometries distributed in a small energy window. Experimental spectra are the result of the Boltzmann statistical average of all energetically favoured structures, which, if some internal rotations are hindered, as it seems to be the case in alanine amorphous film for example \cite{Meinert2015oleb}, can be a subset only of all low energy nuclear geometries of the isolated system. Nevertheless, to simulate such spectra, theoretical calculations should sample a large conformational space and locate all relevant structures corresponding to local minima of the potential energy surface (PES). Finally, amino acids and sugars have plenty of bright states in the ultraviolet (UV), including in the VUV region, while it is not easy to calculate precisely, many high-lying excited states at the same time.

A popular workhorse to calculate the molecular excited states and ECD spectra is time-dependent density functional theory (TDDFT).\cite{Gross1984, Casida1995, Bour1999, Pecul2004} Although TDDFT does not provide multi-excited states and the fact that predicted, mono-excited energy levels may vary depending on the selected exchange-correlation (XC) functional, fairly satisfactory ECD spectra have been obtained in previous studies.\cite{mang2012electronic, nugroho2014circular}  In this article,  the anisotropy spectra $g(\lambda)$ of two chiral molecules, namely valine and glyceraldehyde, will be investigated. A previous theoretical work on the ECD spectra of the valine zwitterion has produced  a theoretical spectra in reasonably good agreement with experiment, with the help of an artificial blue-shift of the energy levels.\cite{ECDValin2009,Tanaka2010}  We will show that TDDFT can produce reasonable anisotropy spectra for valine amorphous film without blue-shift, provided that the right functional is used.  Regarding  glyceraldehyde, experimental synchrotron absorption and ECD spectra have been measured from 330 nm to 180 nm for an amorphous film on \ce{MgF2} window. These spectra have not been published yet. However, we will present the first theoretical calculation of EA, ECD and anisotropy spectra of this system with the same functional validated for valine. Recalling that glyceraldehyde formation has been detected in interstellar ice analog experiments\cite{Meinert2015}, this anisotropy spectra might shed light on the origin of sugar homochirality in nucleic acids.

The article is organized as follows, we present the details of our calculation in section 2. In section 3, we will first discuss the conformers, EA and ECD spectra of the valine zwitterion and glyceraldehyde, obtained from different theoretical calculations. Then, the anisotropy spectra of the two species will be analysed. Section 4 will give a brief conclusion. 

\section{Computational methods}

\begin{figure}
 \begin{tabular}{cc}
	\resizebox{0.3\textwidth}{!}{\includegraphics{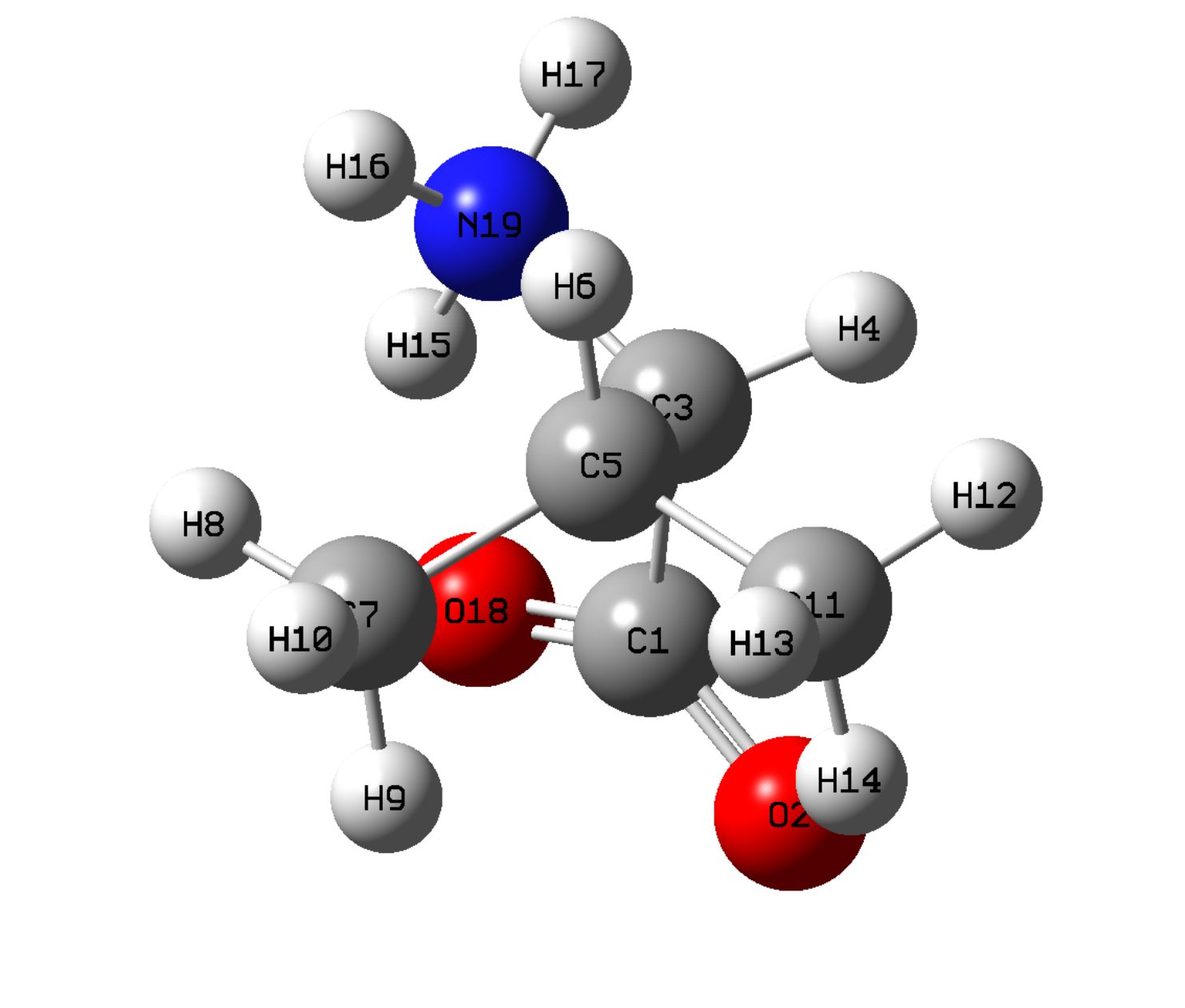}} &
	\resizebox{0.3\textwidth}{!}{\includegraphics{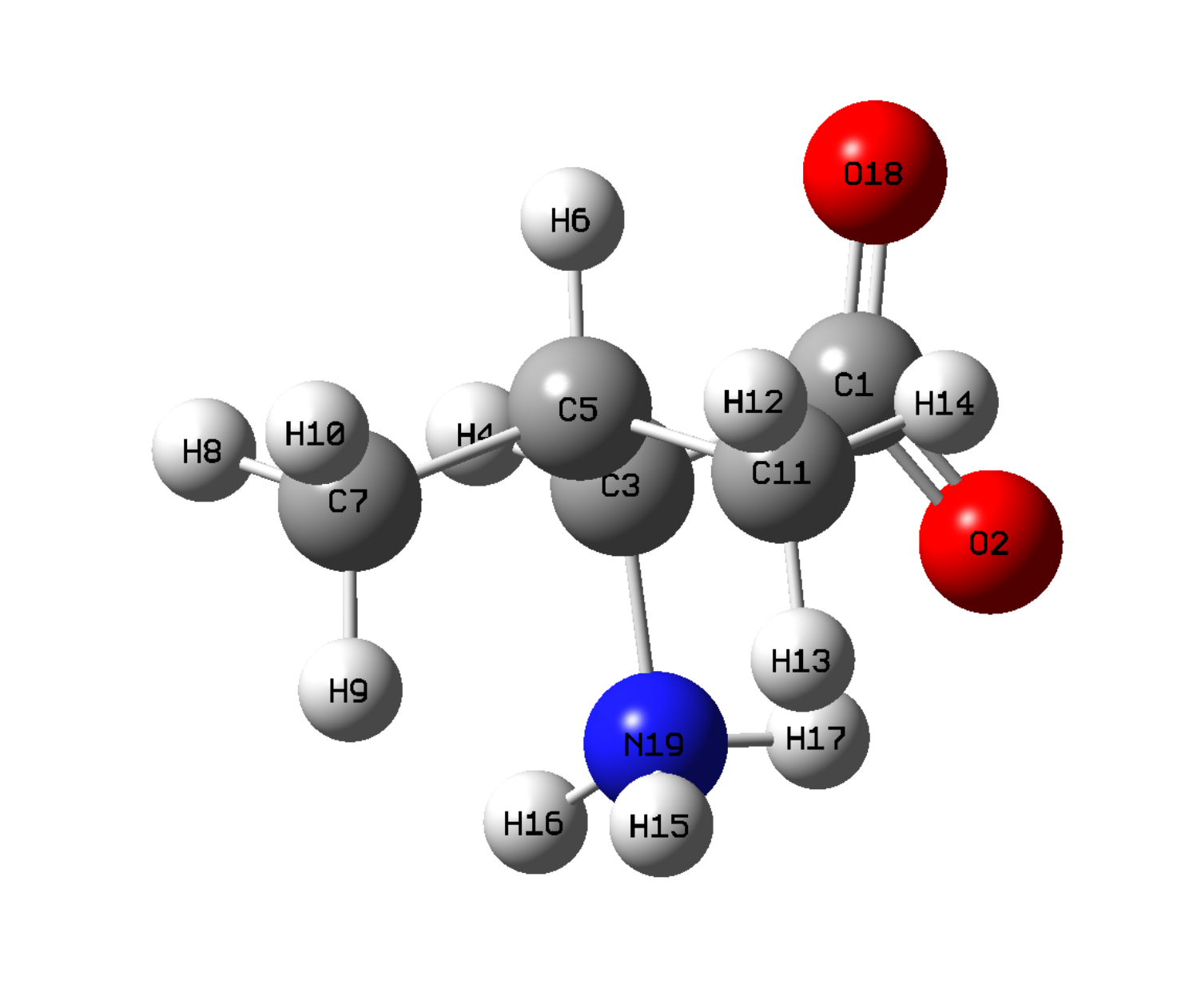}}    \\
	(a) & (b) \\ 
	\resizebox{0.3\textwidth}{!}{\includegraphics{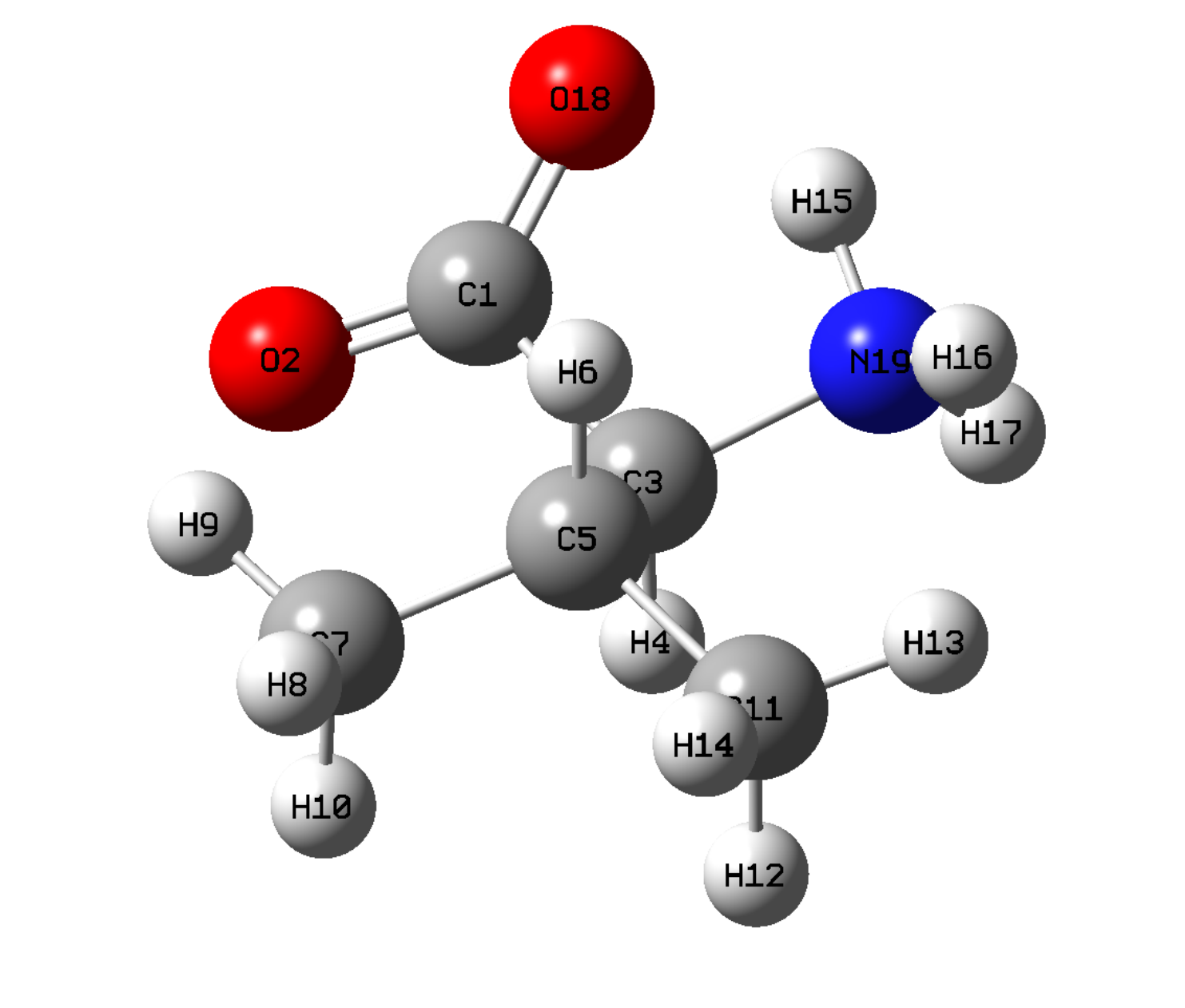}}	&
	\resizebox{0.3\textwidth}{!}{\includegraphics{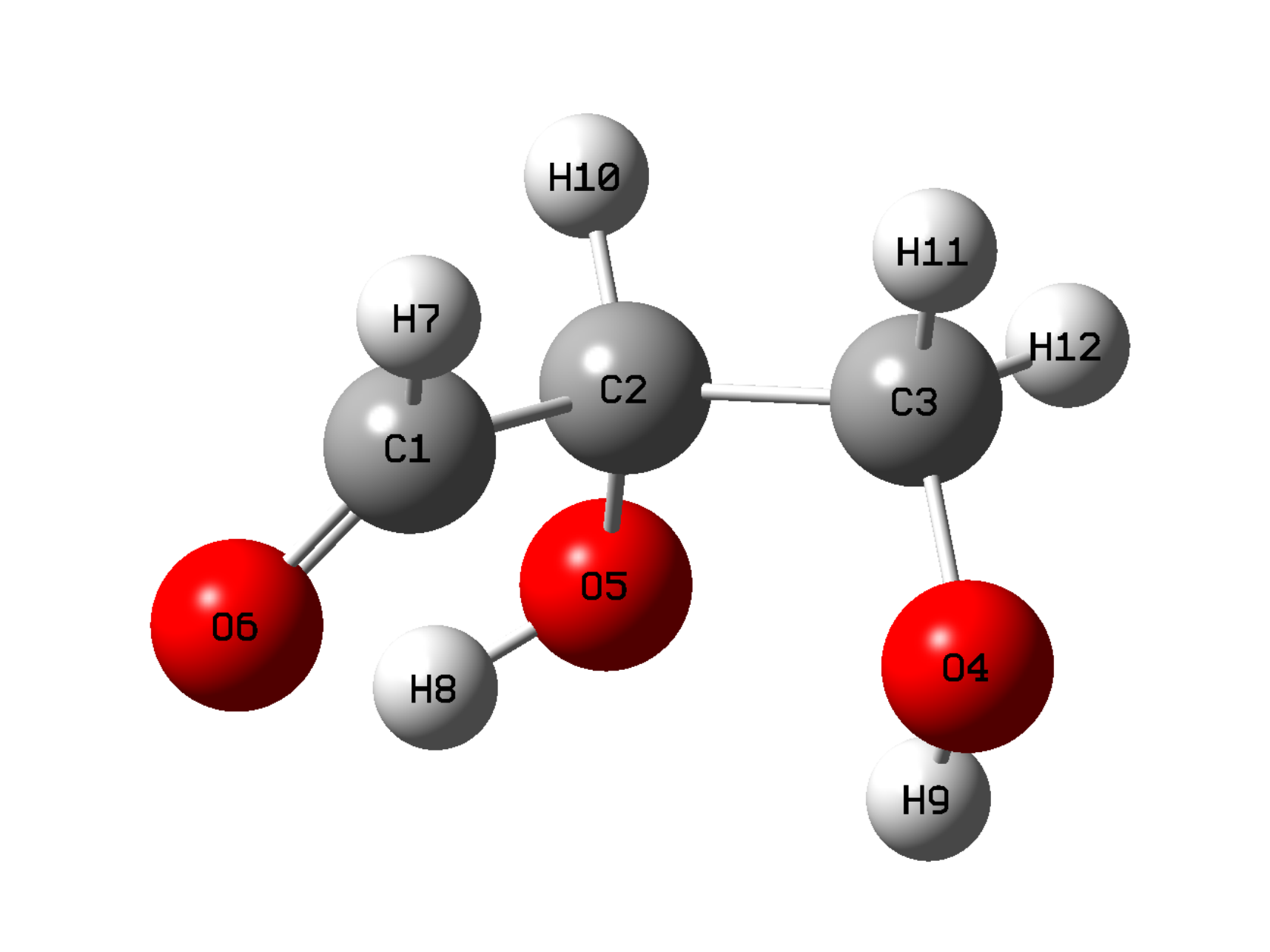}}    	       \\	
	(c) & (d) \\
 \end{tabular}
 \caption{Three Optimized structures of the L-Valine zwitterion and the lowest energy geometry of D-glyceraldehyde. (a) \textit{Trans/Gauche+}; (b) \textit{Gauche+/Gauche-}; (c) \textit{Gauche-/Trans}; (d) D-glyceraldehyde}
\label{fig:Valin-Geom}
\end{figure}

The experimental EA and ECD spectra are the results of a statistical average of energetically allowed conformations. Therefore, a conformational search was performed for valine and glyceraldehyde. The valine molecule exists in the form of a zwitterion in the studied, amorphous films\cite{meierhenrich2010circular} i.e. the proton from the carboxy group in the neutral form is transferred to the amino group. Direct optimization of the zwitterionic structure always results in the proton transferring back to the carboxy group. Therefore, the three N-H bond lengths of the amino group were constrained to 1.01 \AA \ and only the other internal coordinates were relaxed in geometry optimizations. Enforcing bond length constraints is an ad hoc procedure to simulate valine in amorphous film, whose main advantage is its simplicity.  Three optimized structures labelled as \textit{Trans/Gauche+}, \textit{Gauche+/Gauche-} and \textit{Gauche-/Trans}, drawn in Figures \ref{fig:Valin-Geom}(a),  \ref{fig:Valin-Geom}(b), \ref{fig:Valin-Geom}(c), respectively, were obtained for the L-valine zwitterion by density functional theory (DFT) with M06-2X functional\cite{M06} and cc-pVTZ basis set.\cite{dunnning1989} The Cartesian coordinates of the optimized geometries are presented in Tables S1 and S2 of supporting information (SI). We have also carried out polarizable continuum model (PCM) calculation \cite{PCM1981} for comparison with our constrained optimization results and with crystal structures from X-ray diffraction. The PCM optimized structures are reported in Tables S3 and S4.  For D-(+)-glyceraldehyde, that is (2R)-2,3-dihydroxypropanal, DFT with  B3LYP functional\cite{B3LYP}  and cc-pVTZ basis, allowed us to locate 32 equilibrium structures by starting the optimization from different initial geometries obtained by rotating  by intervals of 10 degrees the dihedral C1C2C3O1 and O5C1C2C3 angles (see Figure \ref{fig:Valin-Geom}(d) ). The geometries were reoptimized at M06-2X/cc-pVTZ level. Frequency analyses were performed  for all these structures to check that there were no imaginary frequencies. All geometries are presented in Tables S5 to S11 in SI.

TDDFT calculations  were carried out to calculate the absorption and ECD spectra of the target species. For estimating the influence of basis sets on excitation energy, larger basis sets, namely: cc-pVQZ, aug-cc-pVTZ and aug-cc-pVQZ were used with TDDFT/B3LYP and TDDFT/M06-2X to calculate the first 30 excited states of glyceraldehyde at the lowest energy geometry obtained with the B3LYP functional. The excitation wavelengths and oscillator strengths are listed in Tables S12 and S13 in SI, respectively. As seen in these tables, the excitation energies of the first three excited states are not sensitive to the level of basis sets used, as indicated by the small red-shift of less than 2 nm between the cc-pVTZ and aug-cc-pvQZ results (except for the third energy level with the M06-2X functional, where it reaches 5 nm). However, the excitation wavelengths of other states are systematically reduced by more than 10 nm when diffuse basis functions are added, while extending basis sets from $3$ to $4$ sets of valence orbitals  does not change  significantly the energy differences. This is observed for both functionals. Therefore, aug-cc-pVTZ basis sets are adopted for glyceraldehyde to calculate the excitation energies. Note that the importance of diffuse functions was also observed in Ref. 15.
In contrast, we retain the cc-pVTZ rather than the aug-cc-pVTZ basis set for the valine zwitterion for reasons that will be explained in the next section.

For assessing the XC functional dependence of the excitation energies, five XC functionals B3LYP,\cite{B3LYP} PBE0,\cite{PBE01999} CAM-B3LYP,\cite{CAMB3LYP} $\omega$B97X-D,\cite{wB97XD} and M06-2X\cite{M06} ordered by increasing amount of the Hartree-Fock  (HF) exchange, were used to calculate $30$ excited states of glyceraldehyde at the lowest energy geometry optimized from B3LYP with cc-pVTZ basis sets.  Excitation energies and oscillator strengths are presented in Table S14 in SI. The absorption and ECD spectra of all five functionals are drawn in Figure S2 of SI.  The overall trend is that the excitation wavelengths are blue-shifted when adding more HF exchange with the exception of the first excited state. The M06-2X yields the longest excitation wavelength of 289 nm for the S1 state while the results of PBE0, CAM-B3LYP and  $\omega$B97X-D  are close to 270 nm.  It is noteworthy that a preliminary experimental study has recorded a weak absorption near 290 nm, with no other bump above 195 nm \cite{Meinert-pc}. This is only compatible with  the M06-2X result (which also gives a reasonable IR spectra for glyceraldehyde, as presented in Figure S1 of supporting information).
However, we will also present B3LYP results for sake of comparison. 

For each equilibrium structure, the TDDFT dipole and rotational strength peaks were broadened by Gaussian convolution with full widths at half maximum (FWHM) of 12 nm and 10 nm for absorption and ECD spectra, respectively.  Then, to simulate the experimental conditions, in the absence of precise structural information for the molecules in amorphous films, all these spectra were averaged according to the Boltzmann weights of the corresponding structures,
\begin{equation}\label{eqboltzman}
w_i=\frac{e^{-\varepsilon_i/k_B T}}{\sum  e^{-\varepsilon_i/k_B T}}
\end{equation}
where $\varepsilon_i$  is the energy difference of the i$^{th}$ geometry with the lowest one, $k_B$ is the Boltzmann constant and $T$ is the temperature whose value is set to 300 Kelvin. Notice that we use total energy to calculate the Boltzmann weight rather than the Gibbs free energy, and that zero point energy corrections are neglected.
All calculations have been performed with the Gaussian 09 package.\cite{frisch2009gaussian}

\section{Results and discussion }

\subsection{Structures, absorption and ECD spectra of the L-valine zwitterion}

Table \ref{tbl:valingeom} presents for  different geometries of the valine zwitterion, the relative total energies, the excitation energy of the first excited states, three dihedral angles and their types according to the conformation classification of  Ref. \cite{gorbitz2006structures}
(labelling that is based on the last two dihedral angles). The first three rows correspond to the optimized structures of Figure \ref{fig:Valin-Geom} (a)-(c). The structures of the 4$^{th}$ and 5$^{th}$ rows correspond to the geometries of the first and third rows, respectively, except for the dihedral angles which are those of the related X-ray structures.\cite{gorbitz2006structures}. The structures of the 6$^{th}$ and 7$^{th}$ rows are interpolated between the structures $1$ and $4$, while those of row $8$ and $9$ are interpolated between $3$ and~$5$. 
Finally, three equilibrium geometries, rows 10$^{th}$ to 12$^{th}$ have been optimized with no geometric constraint but with water solvent effect modelled via PCM\cite{PCM1981}.  Harmonic vibrational analysis indicates that these three structures are true minima of the DFT/PCM potential energy surface.

As seen in Figure \ref{fig:Valin-Geom}(a)-\ref{fig:Valin-Geom}(c) and Table \ref{tbl:valingeom}, the dihedrals N19C3C1O2 in all  three geometries from the constrained optimization are small due to the strong hydrogen bond between \ce{COO-} and \ce{NH3+} hindering rotation of \ce{COO-}.  When the valine zwitterion is stabilized by the external polarized field,  intra-molecular hydrogen bond can be weakened, as seen from the O18-H9 distance of 1.737 $\AA$ in the 10$^{th}$ structure much longer than that of 1.525 $\AA$ in the 1$^{st}$ (depicted in Figure \ref{fig:Valin-Geom}(a)). Interestingly, the  three dihedral angles of the 10$^{th}$ are in between the corresponding parameters of the structures from constrained optimization and crystal diffraction. Due to different inter-molecular interactions, these dihedral angles of the valine zwitterion may differ  in the amorphous film from those in crystal or in water solvent.  To simulate these possible geometrical fluctuation, linear interpolations of the three major dihedral angles were performed between the optimized geometries (1$^{st}$ and 3$^{rd}$) and their counterparts in the valine crystal (4$^{th}$ and 5$^{th}$), giving  the four structures ($6^{th}$-$9^{th}$ rows) listed in Table \ref{tbl:valingeom}. Then the absorption, ECD, and anisotropy spectra of these nine structures were averaged 
with Boltzmann weight according to Eq. (\ref{eqboltzman}). The three DFT/PCM optimized geometries (10$^{th}$ to 12$^{th}$) were also used to perform TDDFT/PCM calculations of EA, ECD and g-factor spectra. The Boltzmann averaged curves so-obtained, will be labelled as "M06-2X/PCM''.

\begin{table}[htbp]
\begin{threeparttable}
\caption{Relative total energy $\Delta E$, excitation wavelength $\lambda$ of the first excited state, three main dihedrals, and the structural type of  L-valine zwitterion conformers.\tnote{a}}	\centering
\begin{footnotesize}
\begin{tabular}{ccccccc}
\hline
\hline
$No.$ & $\Delta E$ (KJ/mol) & $\lambda(nm)$ &  N19C3C1O18                                                           ($^\circ$) &\begin{footnotesize} N19C3C5C11                                                                                                                                                                                    \end{footnotesize} ($^\circ$) &  \begin{footnotesize}N19C3C5C7                                                                                                                                                                                                                                                                                                                                                                                                                                      \end{footnotesize} ($^\circ$)  & Type\\
\hline
1  &   0.00     & 211  &  -5.8   &  -168.2  &  68.3 & \textit{Trans/Gauche+}    \\
2  &   0.05     & 211  &  1.0    &  60.3    & -66.1    &   \textit{Gauche+/Gauche-}  \\
3  &   10.39   & 212 &  -11.4  &  -61.8   & 175.7 & \textit{Gauche-/Trans}    \\
4  &   8.14     & 214  &  -22.2  &  -154.4  &  81.1  & \textit{Trans/Gauche+}   \\
5  &   19.53    & 221  &  -37.7  &  -52.4   & -175.6 & \textit{Gauche-/Trans}  \\
6   &   4.99     & 213  &  -11.1  &  -163.6  & 72.6  & \textit{Trans/Gauche+}  \\
7 &   19.79    & 214  &  -16.7  &  -159.0  & 76.8  & \textit{Trans/Gauche+}  \\
8  &   12.08    & 216  &  -20.2  &  -55.5   & 178.9 & \textit{Gauche-/Trans} \\ 
9 &   14.18    & 216  &  -28.9  &  -58.8   & -179.0 & \textit{Gauche-/Trans} \\
10  &   7.67     & 216  &  -13.3  &  -167.3  & 68.3  & \textit{Trans/Gauche+}    \\
11 &   4.16      & 215  &  -3.6   &  61.8       & -64.4 & \textit{Gauche+/Gauche-}   \\
12 &   15.28   & 219  &  -22.7  &  -67.7   & 169.7 & \textit{Gauche-/Trans}     \\
\hline 
\end{tabular}
\label{tbl:valingeom}
\end{footnotesize}
\begin{tablenotes}
  \item[a]  The dihedral angles of structures 1-3 were obtained from geometry optimization by constraining the NH bond to 1.01 \AA; the dihedral angles of structures 4 and 5 are taken  from the Ref. \cite{gorbitz2006structures}; structures 6 and 7 correspond to linear interpolations between 1 and 4; 8 and 9  to linear interpolations  between 3 and 5; structures 10-12 are from DFT optimization with PCM for water solvent
\end{tablenotes}
\end{threeparttable}
\end{table}

\begin{figure}
\begin{tabular}{cc}
    \resizebox{0.5\textwidth}{!}{\includegraphics{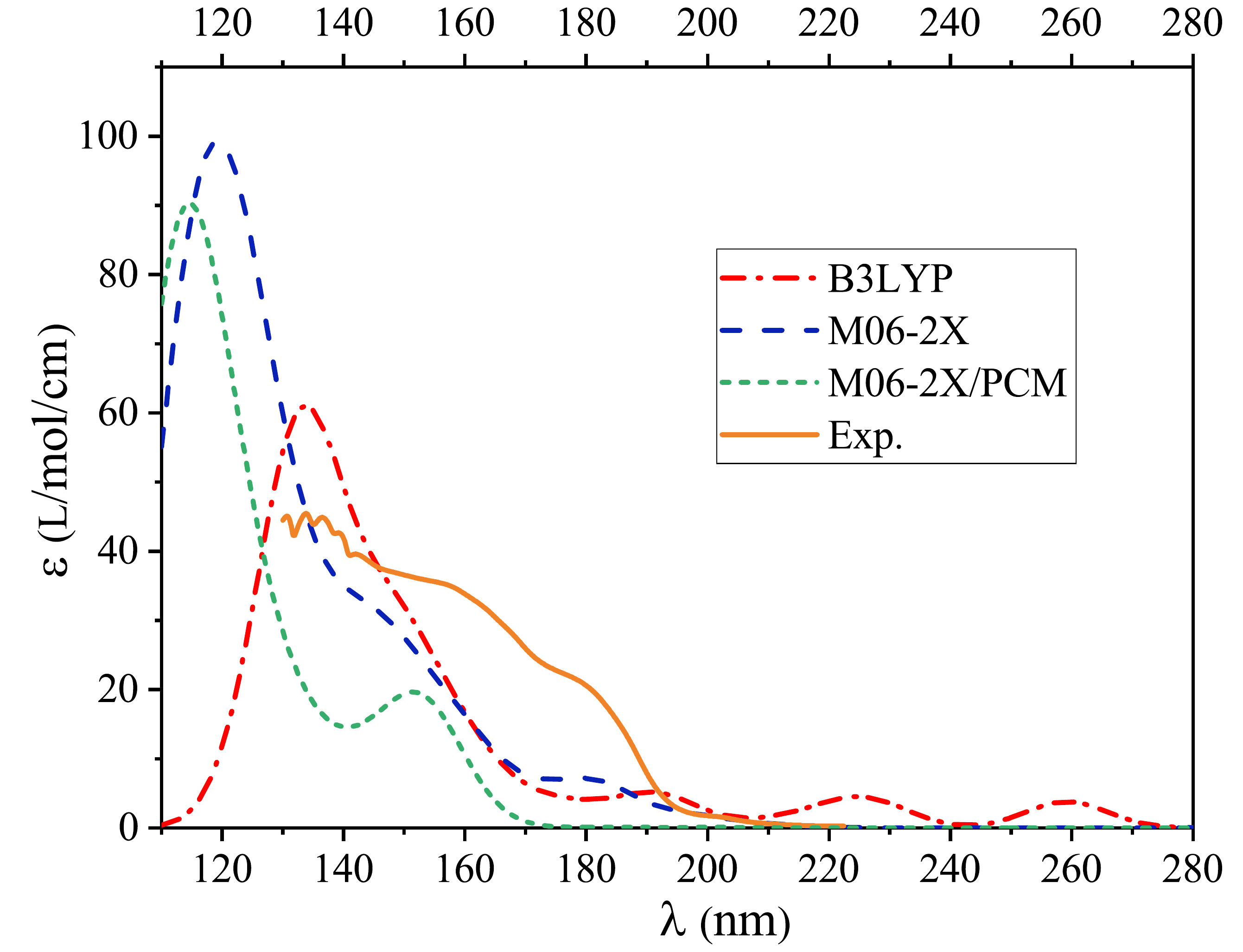}} &
     \resizebox{0.5\textwidth}{!}{\includegraphics{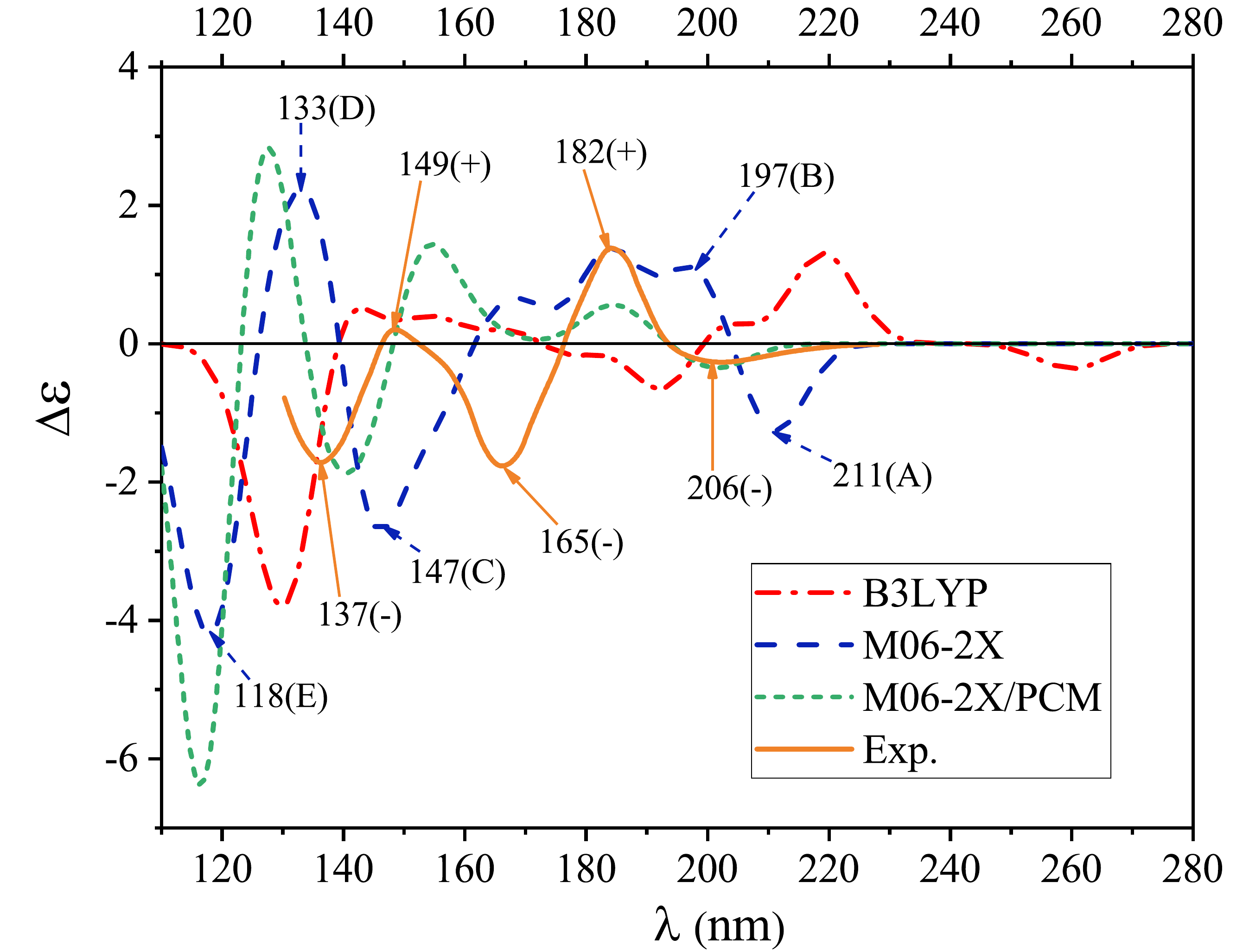}}  \\
     (a) & (b)
\end{tabular}
\caption{ Absorption and ECD spectra of the L-valine zwitterion: (a) absorption spectra (b) ECD spectra. The EA and ECD  experimental spectra are redrawn from Ref. 30 and Ref. 25, respectively.}
\label{fig:Valine-spec}
\end{figure}

The Boltzmann weighted EA and ECD spectra, as well as the experimental spectra\cite{Bredehoft2014gsolvent, Meinert2012anisotropy} of the L-valine zwitterion are drawn in Figure \ref{fig:Valine-spec}(a) and \ref{fig:Valine-spec}(b), respectively. The excitation wavelengths, oscillator strengths, and rotatory strengths  of the first 20 excited states of the lowest-energy geometry are listed in Table \ref{tbl:valstates} for both M06-2X and B3LYP functionals, for comparison.

Generally speaking, the absorption spectra from TDDFT/B3LYP has a noticeable red-shift compared to the M06-2X one, as seen in Figure \ref{fig:Valine-spec}(a). As reported in Table \ref{tbl:valstates}, the first excited state from the TDDFT/M06-2X calculation appears at 211 nm with a small oscillator strength of 0.0016. The S2 state has an excitation wavelength of 197 nm and larger oscillator strength of 0.0081, indicating a stronger absorption than that of S1. In contrast, the first two excited states from TDDFT/B3LYP are located at 257 nm and  225 nm, and the order of their oscillator strengths (0.1770 and 0.0013,  respectively), is reversed. In the experimental spectra, the valine film starts to absorb the UV light at about 215 nm, and has a sharp rise at 195 nm, suggesting that the absorption of S1 is weaker than S2.\cite{ECDValin2009, Meinert2012anisotropy} Therefore, TDDFT/M06-2X produces more reasonable results than TDDFT/B3LYP taking into account the excitation energies, the oscillator strengths, and also the energy gap between S1 and S2. 

\begin{figure}
\begin{tabular}{c}
    \resizebox{0.8\textwidth}{!}{\includegraphics{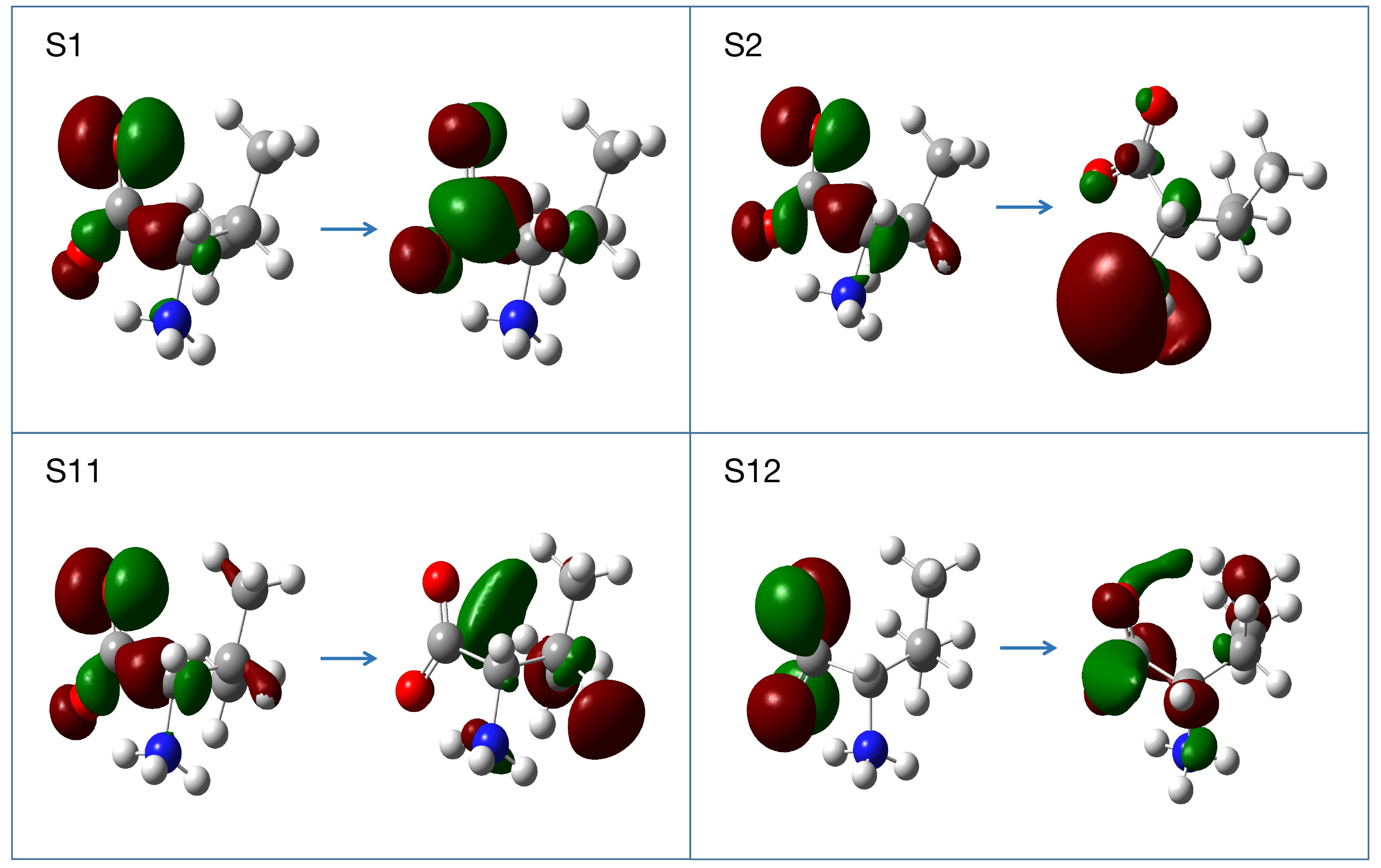}} 
\end{tabular}
\caption{Natural transition orbitals (NTOs) of the S1, S2, S11 and S12 states of the valine zwitterion from TDDFT/M06-2X. Left are the hole NTOs, right the particle NTOs.}
\label{fig:Valine-NTO}
\end{figure}

Natural transition orbital (NTO)  analysis from TDDFT/M06-2X (see in Figure \ref{fig:Valine-NTO}) shows that the S1 state arises from a $n\rightarrow\pi^*$ transition, where $n$ is a C1-C3 bonding orbital mixed with lone pair orbitals from two oxygen atoms, and the $\pi^*$ orbital is an anti-bonding one between C and O in \ce{COO-}. The hole NTO of S2 is the same as that of S1 but the particle NTO mainly distributes on the \ce{NH3+} group. Interestingly, NTOs from TDDFT/B3LYP indicate that the S1 and S2 states are exchanged. TDDFT/M06-2X using aug-cc-pVTZ basis sets also give the inverse order for S1 and S2 states like TDDFT/B3LYP. Thus, we adopt the cc-pVTZ  basis sets to calculate the valine zwitterion.

The absorption of the valine film has a sharp rise from 185 nm to 170 nm (see supporting information of Ref. \cite{Meinert2012anisotropy}), giving rise to a shoulder structure on the experimental spectrum. In agreement with this, we could identify a shoulder structure near 180 nm in the theoretical spectrum without PCM, that arises from the S3 and S4 states at 181 nm and 178 nm. These states have much larger oscillator strengths than those of the nearby states, as demonstrated in Table \ref{tbl:valstates}. Furthermore, we can locate another shoulder from 170 nm to 145 nm on the theoretical spectra, which should correspond to the shoulder from 170 to 155 nm in the experimental work. As can be seen in Table \ref{tbl:valstates}, several excited states, from S6 to S19, are distributed in this region. All of them have substantial oscillator strengths. Overall, TDDFT/M06-2X produces an EA spectrum that agrees well with the experimentally measured one. 

%

\begin{table}[htbp]
\caption{Comparison of the excitation wavelength $\lambda$, dipole oscillator strength $f$, and rotatory strength $R$ of the first 20 excited states of the L-valine zwitterion  calculated by TDDFT/M06-2X and TDDFT/B3LYP at the lowest-energy structure.}  \centering\
\label{tbl:valstates}
\begin{tabular}{cccccccc}
  \hline
  \hline
  \multirow{2}{*}{State} & \multicolumn{3}{c}{M06-2X}  & \multicolumn{3}{c}{B3LYP}\\
  \cline{2-4}\cline{6-8}
    & $\lambda$(nm) & $f$   & $R$   & & $\lambda$(nm) & $f$   & $R$ \\
   \hline
1    &    211 &  0.0012 &   -12.2511 &  & 257  &  0.0177  &  -6.6521   \\
2    &    197 &  0.0081 &   13.9018  &  & 225  &  0.0013  &  -1.4187   \\
3    &    181 &  0.0130 &   18.4377  &  & 223  &  0.0120  &  4.1660   \\
4    &    180 &  0.0016 &   3.0047   &  & 218  &  0.0070  &  25.8166   \\
5    &    178 &  0.0113 &   -4.9941  &  & 209  &  0.0034  &  -3.3600   \\
6    &    169 &  0.0146 &   3.6546   &  & 191  &  0.0037  &  -15.4107   \\
7    &    161 &  0.0176 &   8.4576   &  & 189  &  0.0055  &  3.1835    \\
8    &    157 &  0.0208 &   -7.1196  &  & 188  &  0.0034  &  -0.3084   \\
9    &    154 &  0.0044 &   -13.9811 &  & 183  &  0.0053  &  -0.9284   \\
10   &    151 &  0.0406 &   0.0706   &  & 178  &  0.0074  &  4.0803   \\
11   &    147 &  0.0202 &   -16.1892 &  & 174  &  0.0051  &  -7.2763    \\
12   &    143 &  0.0347 &   -30.342  &  & 166  &  0.0016  &  2.3315   \\
13   &    141 &  0.0040 &   -0.7203  &  & 164  &  0.0069  &  6.3020   \\
14   &    139 &  0.0047 &   -9.8122  &  & 160  &  0.0172  &  -5.8125   \\
15   &    138 &  0.0165 &   27.8018  &  & 159  &  0.0078  &  8.6033   \\
16   &    136 &  0.0051 &   6.4944   &  & 156  &  0.0012  &  7.1900  \\
17   &    136 &  0.0115 &  -5.5526   &  & 153  &  0.0059  &  5.5724   \\
18   &    136 &  0.0075 &   14.6826  &  & 152  &  0.0124  &  -8.9647   \\
19   &    135 &  0.0143 &   12.1560  &  & 151  &  0.0046  &  1.8114   \\
20   &    133 &  0.0119 &   -10.1010 &  & 150  &  0.0088  &  -3.0247  \\
  \hline
\end{tabular}
\end{table}

The longest wavelength peak of the ECD spectra will be labelled as peak A. The ECD spectra corresponding to the individual geometries of Table \ref{tbl:valingeom} have been  plotted in Figure S4 of SI. Figure S4(c) gathers the geometries calculated with PCM, while the others are displayed in two separate frames according to the sign of peak A: Spectra with positive peak A are gathered in Figure S4(b) and those with negative peak A in Figure S4(a). As can be seen in Figure S4, all \textit{Trans/Gauche+} geometries have a positive peak near 215 nm except for  the  4$^{th}$ structure in Table \ref{tbl:valingeom}, which is extracted from crystal X-ray diffraction.  This structure has the largest N19C3C1O18 angle (-22.2$^{\circ}$) from all  \textit{Trans/Gauche+}  geometries. Regarding the \textit{Gauche-/Trans} structures, the two DFT-optimized  ones ($3^{rd}$ and $12^{th}$ in Table \ref{tbl:valingeom}) show  a  negative peak near 215 nm, while all the others, including the structure from crystal diffraction, have a positive peak in this region. In general, the amplitudes of the negative A peaks  are higher than those of the positive ones. As a result, a negative peak is produced in the long wavelength region, when all geometries are averaged.  The Gauche+/Gauche- structure from the constrained optimization has a negative A peak, so has the equilibrium structure in water solvent. The Gauche+/Gauche- structure has not been reported in valine crystals, whereas our calculations indicate that it should be lower in energy than the Gauche-/Trans observed one.

Now let us compare the Boltzmann weighted ECD spectra with the experimental result measured in the valine film. The experimental spectra are characterized by five peaks at 206(-) nm, 182(+) nm, 165(-) nm, 149(+) nm and 137(-) nm, 
as seen in Figure \ref{fig:Valine-spec}(b)\cite{ECDValin2009, Tanaka2010, meierhenrich2010circular}.  The peak at 206(-) nm should arise from the S1 state, as supported by the small negative peak A centered at 211 nm in the M06-2X theoretical calculation. The peak centered at 182(+) nm has a smooth shape and spreads from 195 to 175 nm. The shape of the M06-2X theoretical spectrum between 206 to 162 nm is irregular, and three peaks at 197 nm, 185 nm and 165 nm with positive signs can be identified in this region. The peak at 197 nm arises from the S2 state and will be referred to as peak B thereafter. We assign it to the experimental peak at 182 nm. The experimental spectrum has a large negative peak at 165 nm.  On the theoretical side, the second negative  M06-2X  peak is centered at  147 nm and will be referred to as peak C. We assign peak C to the  165 nm  peak because the latter only appears on the experimental spectra of amino acid with an alkyl group in the side chain.\cite{Tanaka2010} As Table \ref{tbl:valstates} shows, the S11 and S12 states at 146 nm and 143 nm are the states with a large rotatory strength that are responsible for peak C.  The hole NTOs of these two states largely spread on the \ce{CH3} group (Figure  \ref{fig:Valine-NTO}), supporting speculations from experimental studies.\cite{Tanaka2010}   The last two experimental peaks at 149(-) nm and 137(+) nm also shift to the blue region at 133(-) ("peak D'') and 118(+) nm ("peak E'') in the theoretical calculation.

The amorphous film ECD spectrum is certainly influenced by the electrostatic field induced by valine zwitterions. However, it is a tremendous task to simulate such an effect, even by using a simple point charge model in the framework of QM/MM, because the precise structure of the amorphous film is totally unknown. To estimate the effect of an external field to the ECD spectrum, we have optimized geometries 10 to 12 in Table \ref{tbl:valingeom} by using the PCM for water solvent and we have Boltzmann averaged the corresponding ECD spectra. The result is also presented in Figure \ref{fig:Valine-spec}(b).  Compared to the spectrum without PCM, all peaks have a systematic blue shift. Peaks A and B appear at 207(-) nm  and 184(+) nm respectively, agreeing fairly well with the experimental result for the amorphous film. In contrast, the other three peaks (C, D and E) being even more blue-shifted than those calculated without PCM, the agreement with experiment is worsened. The wavy region between peaks B and C, has only one extra positive peak with PCM instead of two. However, whether PCM is used or not, the description of this region is not satisfactory. 

Tanaka \textit{et. al} had calculated the ECD spectra of valine film by averaging the spectra of two structures obtained from a crystal structure database.\cite{Tanaka2010}  A small basis sets 6-31+G(d,p) and the B3LYP functional were used in their calculation. Their theoretical spectrum had been blue-shifted by 0.6 eV to adjust calculated ECD peaks to the experimental ones. In contrast, the ECD spectra in this work are obtained with no artificial shift, for a match to experiment by no means inferior to Tanaka's results.

\subsection{Structures, absorption and ECD spectra of glyceraldehyde}

\begin{figure}
 \begin{tabular}{cc}
	\resizebox{0.5\textwidth}{!}{\includegraphics{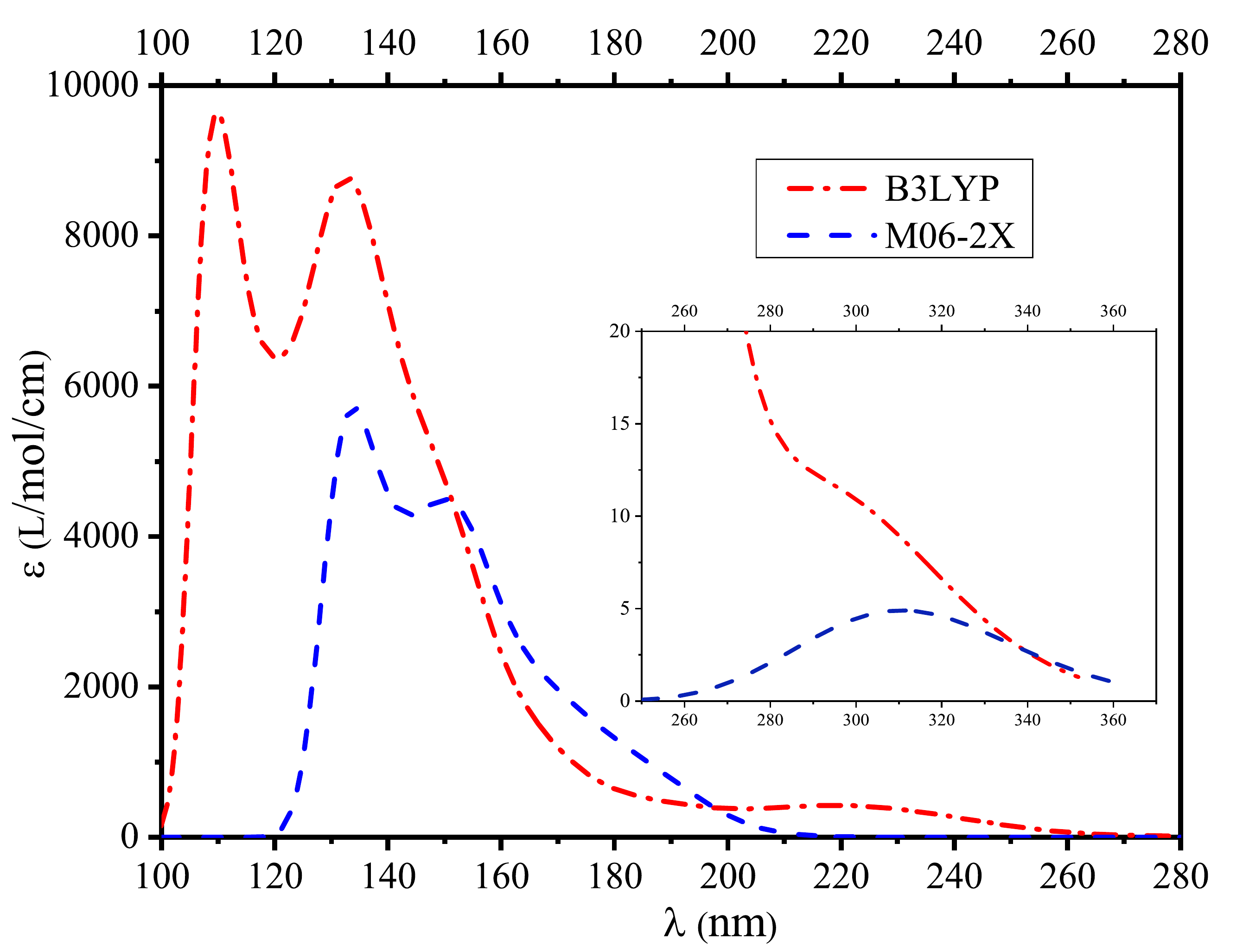}} &
	\resizebox{0.5\textwidth}{!}{\includegraphics{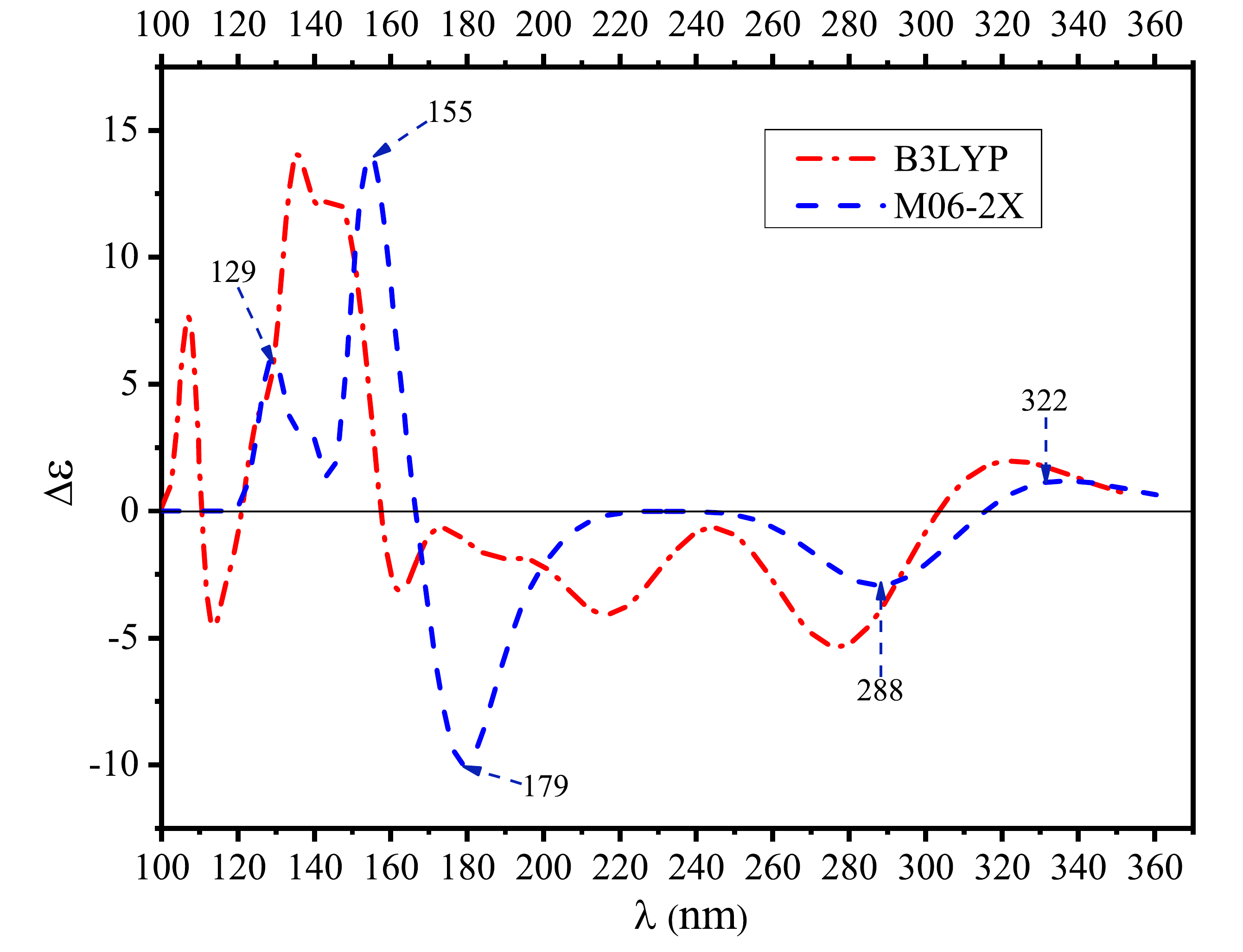}} \\
	(a) & (b)  \\
     \end{tabular}
	\caption{TDDFT absorption and ECD spectra of D-glyceraldehyde obtained with the aug-cc-pVTZ basis set, by using the M06-2X and B3LYP functionals. (a) EA spectra. (b) ECD spectra. }
	\label{fig:Gly-Spectra}
\end{figure}

Table \ref{tbl:glygeom} lists the relative total energy, excitation wavelength, dipole oscillator strength, rotatory strength of the first excited state, and two dihedral angles of 32 optimized structures of D-glyceraldehyde calculated by TDDFT/M06-2X with aug-cc-pVTZ basis sets. The Boltzmann weighted absorption and ECD spectra of D-glyceraldehyde from TDDFT with M06-2X and  B3LYP functionals are presented in Figure \ref{fig:Gly-Spectra}.  The excitation wavelength $\lambda$, dipole oscillator strength $f$,  rotatory strength $R$ of the first 20 excited states of the lowest-energy geometry of D-glyceraldehyde calculated by TDDFT with two different functionals  for comparison, are presented in Table \ref{tbl:Gly-ES}. 

\begin{table}[htbp]
\caption{ Relative total energy $\Delta E$, excitation wavelength $\lambda$ of the first excited state with the corresponding dipole oscillator strength $f$ and rotatory strength $R$, and two key dihedral angles of 32 conformers of D-glyceraldehyde obtained  at M06-2X/aug-cc-pVTZ  TDDFT level.}	\ \centering\
\label{tbl:glygeom}
\begin{tabular}{cccccccc}
		\hline
		\hline
		$No. $ & $\Delta E$ (KJ/mol) & $\lambda$(nm)  & f & $R$ &   O6C1C2C3 ($^\circ$) &  C1C2C3O4 ($^\circ$)  \\
\hline
1  & 0          & 288  &  0.0001 & -6.4499 &-126.3   &   60.3    \\
2  & 4.66   & 293  &  0.0003 &-12.3435 &-118.2   &   178.7   \\
3  & 7.47   & 319  &  0.0003  &11.9961 &8.3      &   64.5    \\
4  & 9.40   & 294  &  0.0001  &-8.2137 &-123     &   -62.9   \\
5  & 10.70  & 294  &  0.0001  &-10.3402 &-124.4   &   -63.1   \\
6  & 12.24  & 315  &  0.0000  &-9.6257 &48.6     &   70.4    \\
7  & 13.46  & 287  &  0.0001  &2.2599 &-125.3   &   50.7    \\
8  & 14.37  & 311  &  0.0001   &-0.3631 &44.1     &   178.3   \\
9  & 15.98  & 309  &  0.0001 &-0.3252 &30.0     &   -68.9   \\
10 & 15.70  & 313  &  0.0001 &3.2308 &59.9     &   174.5   \\
11 & 17.71   & 320  &  0.0002 &4.9252 &16.9     &   -64.8   \\
12 & 17.14   & 315  &  0.0006 &-12.8125 &-91.9    &   68.0    \\
13 & 18.55  & 321  &  0.0001  &6.8206 &70.9     &   -180    \\
14 & 18.92  & 318  &  0.0000 &5.2564 &78.9     &   61.8    \\
15 & 19.82  & 309  &  0.0001 &3.6866 &61.3     &   63.6    \\
16 & 19.21  & 292  &  0.0001  &-6.5259 &-124.7    &   -163.9  \\
17 & 22.10  & 322  &  0.0002 & 3.0277 &30.2     &   -175.7  \\
18 & 21.06  & 294  &  0.0002 &-12.1897 &-123.9   &   -174.1  \\
19 & 27.25  & 303  &  0.0001  &-8.5460 &-123.1   &   60.5    \\
20 & 27.01  & 315  &  0.0002  &7.7219 &23.4     &   53.4    \\
21 & 27.30  & 308  &  0.0001  &-3.4812 &-118.5   &   -177.9  \\ 
22 & 29.65 & 310  &  0.0001  &-5.4518 &-115.6   &   177.8   \\
23 & 29.98 & 320  &  0.0001 &1.2921 &135.0    &   -57.6   \\
24 & 29.90 & 315  &  0.0000 &3.7045 &78.8     &   53.1    \\
25 & 31.89  & 322  &  0.0002 &5.4518 &63.5     &   -170.8  \\
26 & 35.41  & 313  &  0.0002  &6.9185 & 97.4    &   -69.6   \\
27 & 36.62  & 322  &  0.0001  &7.9033 & -20.9   &   -68.4   \\
28 & 36.86  & 311  &  0.0001  &-6.6276 &-125.5   &   -66.3   \\
29 & 37.18  & 317  &  0.0000  &1.8084 & 112.2   &   -58.5   \\
30 & 39.25 & 312  &  0.0001  &-6.8418 &-126.4   &   -62.9   \\
31 & 39.86  & 313  &  0.0001  &1.5423 & 9.2     &   -76.0   \\
32 & 41.74  & 318  &  0.0001  &6.1216 &5.0      &   -75.2   \\
\hline
\end{tabular}
\end{table}

Several authors have previously studied glyceraldehyde conformers both theoretically and experimentally, see Refs. \cite{Lozynski1997,Powis2000,Pecul2002,lovas2003,vogt2009} to quote a few.
As can be seen in Table \ref{tbl:glygeom}, the geometry with the highest total energy is only $41.74$ KJ/mol above the lowest one, indicating that glyceraldehyde is very floppy. Notice that Vogt and co-workers have located $36$ conformers with B3LYP/cc-pVTZ calculations by hunting them on a $5$-dimensional section of the potential energy surface.\cite{vogt2009} Among these conformers, $31$ can be identified in our results. However, their B3LYP/cc-pVTZ energy ordering of the lowest $5$ conformers differs from our M06-2X/aug-cc-pVTZ ordering, in that our third conformer is found only in fifth position. The B3LYP/cc-pVTZ ordering is the same as that found by Lovas et al. at MP2/6-311++G** calculation level (but starting from B3LYP results).\cite{lovas2003}.  In contrast, the  M06-2X/aug-cc-pVTZ ordering of the first five conformers is the same as that of the MP2/cc-pVQZ calculations of Vogt et al. (see supporting information).\cite{vogt2009} Note that in support to the latter ordering, a microwave study has identified our $3$ first conformers but neither the $4^{th}$ nor the  $5^{th}$ conformer that  B3LYP/cc-pVTZ and MP2/6311++G** found lower in energy.

\begin{table}[h]
\caption{Comparison of the excitation wavelength $\lambda$, dipole oscillator strength $f$,  rotatory strength $R$ of the first 20 excited states of D-glyceraldehyde calculated  by TDDFT/M06-2X and TDDFT/B3LYP at the lowest-energy geometry with the aug-cc-pVTZ basis set.} \centering\
\label{tbl:Gly-ES}
\begin{tabular}{cccccccc}
  \hline
  \hline
  \multirow{2}{*}{State} & \multicolumn{3}{c}{M06-2X}  & \multicolumn{3}{c}{B3LYP}\\
  \cline{2-4}\cline{6-8}
    & $\lambda(nm)$ & $f$   & $R$   & & $\lambda(nm)$ & $f$   & $R$ \\
   \hline 
1    &    288 &  0.0001 &   -6.1035   & & 275  &  0.0001  &  -7.9344   \\
2    &    185 &  0.0270 &   -19.8234 &  & 230  &  0.0151  &  -0.7521   \\
3    &    170 &  0.0133 &   3.3057     &  & 204  &  0.0062  &  -13.5072  \\
4    &    167 &  0.0099 &   1.8848    &  & 169  &  0.0062  &  3.6294    \\
5    &    165 &  0.0006 &   -7.2808  &  & 165  &  0.0023  &  12.7276   \\
6    &    163 &  0.0085 &   -15.1908 &  & 163  &  0.0044  &  5.7125    \\
7    &    157 &  0.0713 &   40.6152    &  & 160  &  0.0118  &  3.9061    \\
8    &    155 &  0.0121 &   7.9807      &  & 159  &  0.0087  &  -26.8748  \\
9    &    154 &  0.0089 &   16.3031   &  & 154  &  0.0880  &  22.8567   \\
10   &    149 &  0.0247 &  -17.1176  &  & 149  &  0.0208  &  3.2984    \\
11   &    148 &  0.0112 &   1.5426      &  & 144  &  0.0107  &  0.4992    \\
12   &    147 &  0.0149 &   -13.4758 &  & 143  &  0.0381  &  4.9907    \\
13   &    146 &  0.0112 &   19.3435   &  & 141  &  0.0455  &  13.9929   \\
14   &    142 &  0.0052 &   4.3228    &  & 140  &  0.0244  &  -28.6825  \\
15   &    141 &  0.0034 &   4.7636    &  & 139  &  0.0144  &  3.3023    \\
16   &    140 &  0.0061 &   3.4244    &  & 138  &  0.0051  &  -4.6863   \\
17   &    139 &  0.0054 &   -9.0742  &  & 135  &  0.0436  &  11.8780   \\
18   &    139 &  0.0070 &   0.7750    &  & 134  &  0.0458  &  15.2718   \\
19   &    137 &  0.0302 &   -8.5185   &  & 132  &  0.0357  &  4.8782    \\
20   &    137 &  0.0106 &   2.3694    &  & 131  &  0.0033  &  -18.2336  \\
  \hline
\end{tabular}
\end{table}

\begin{figure}
\begin{tabular}{c}
    \resizebox{0.8\textwidth}{!}{\includegraphics{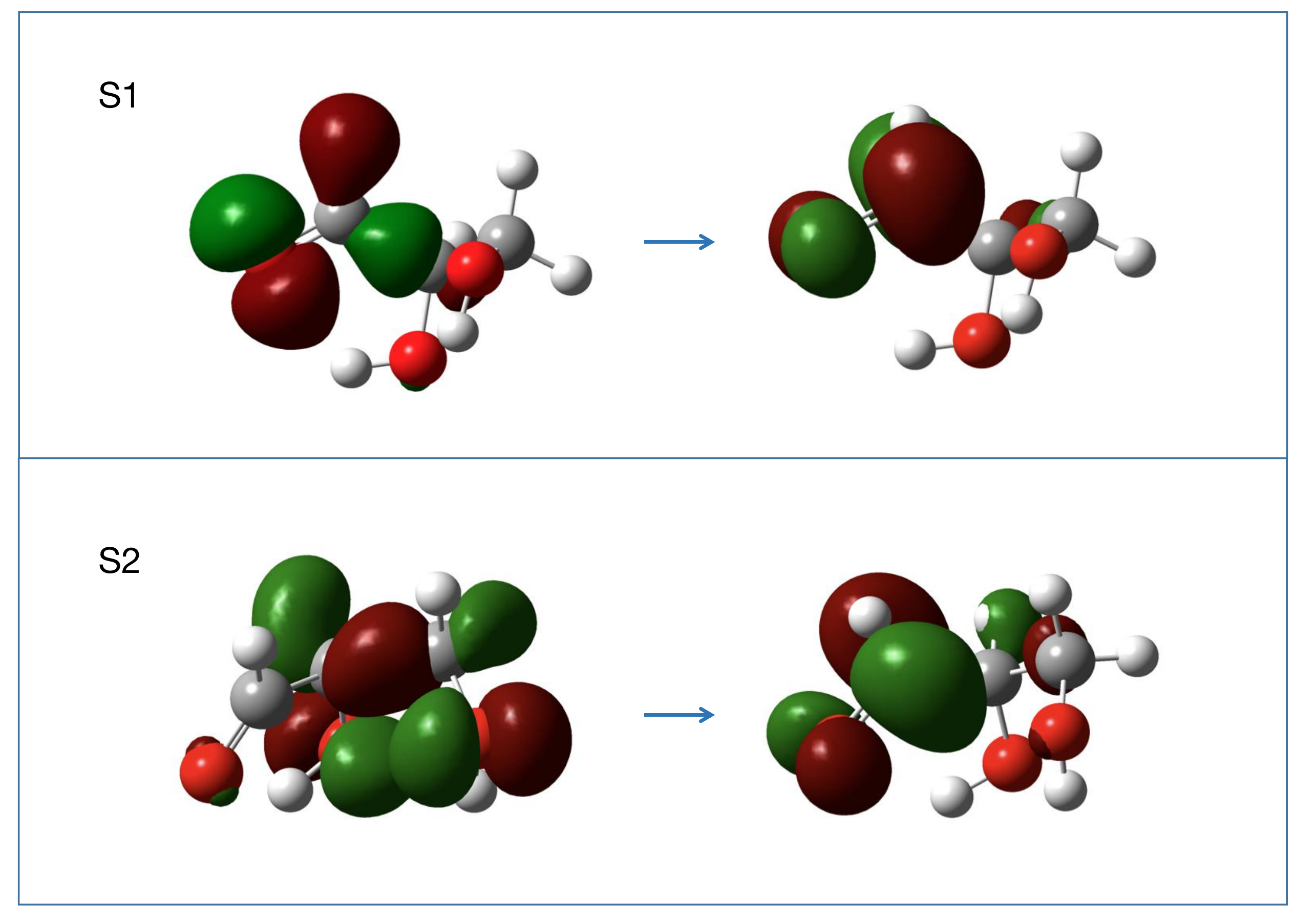}} 
\end{tabular}
\caption{Natural transition orbitals (NTOs) of the S1 and S2 states of glyceraldehyde from TDDFT/M06-2X. Left are the hole NTOs, right the particle NTOs.}
	\label{fig:nto-gly}
\end{figure}

As seen in Table \ref{tbl:glygeom}, the distribution of excitation wavelengths of the S1 state of the 32 tabulated geometries  ranges from $288$ to $322$ nm. Interestingly, the oscillator strength of S1 at the lowest energy geometry is close to zero while some other geometries have larger oscillator strengths, indicating that internal rotation could be responsible for the absorption of glyceraldehyde in this region. These small oscillator strengths give rise to a tiny absorption peak centered at $318$ nm as shown in Figure \ref{fig:Gly-Spectra}(a). The NTOs displayed in Figure \ref{fig:nto-gly} show that the hole NTO of the S1 state is a combination of an O6 lone pair with the C1-C2 $\sigma$-bonding orbital.  The particle NTO is an anti-bonding $\pi^*$ orbital of the C1=O6 double bond. This observation supports our hypothesis that the distortion of the  O6C1C2O5 dihedral angle can promote the absorption of S1.  Another interesting finding is that the  rotatory strengths of S1 for different conformers can have opposite signs,  leading to two moderate opposite peaks centered at 288 nm and 335 nm on the ECD spectra after statistical averaging. More precisely, the third conformer has a positive rotatory strength while the four other conformers in the five lowest ones have negative rotatory strengths. Therefore, ECD spectra may be used to detect if the $3^{rd}$ conformer can exist under given experimental conditions as it was the case in a microwave  study.\cite{lovas2003}  TDDFT/B3LYP  gives a similar feature for the S1 state, but the two peaks are blue-shifted by about  $15$ nm compared to TDDFT/M06-2X. The small dipole strength and large rotatory strength of S1 will give rise to large anisotropic factors in the long wave length region.

The  two XC functionals produce different TDDFT energy gaps between the S1 and S2 states. As seen in Table \ref{tbl:Gly-ES}, with M06-2X, the S2 state is $103$ nm away from S1, while the same gap with B3LYP is only of $45$ nm. The S3 state has an excitation wavelength of $170$ nm (TDDFT/M06-2X), which is blue-shifted by $15$ nm from S2. Both the S2 and S3 states have much larger oscillator strength than that of S1, giving rise to two shoulders starting at $188$ nm and $162$ nm on the M06-2X absorption spectra. The B3LYP EA spectrum is qualitatively different with a bump between $230$ and $210$ nm, and no clear shoulder structure at shorter wavelength. The two XC functionals also produce very distinct ECD features for S2 and S3. As seen in Figure \ref{fig:Gly-Spectra}(b),  there are two high peaks with opposite sign at $180$ nm and $170$ nm in the M06-2X result while TDDFT/B3LYP gives two small negative peaks at $220$ nm and $188$ nm. The hole NTO (Figure \ref{fig:nto-gly}) of S2  is a combination between the  C2-C3 $\sigma$-bonding  orbital  and lone pair orbitals of the O4 and O5 atoms. The particle NTO (Figure \ref{fig:nto-gly}) is dominated by the C1=O6 anti-bonding $\pi^*$ orbital. Therefore, the S2 state arises from an $n\rightarrow\pi^*$ transition. In contrast, the NTOs of the S3 state distribute on many atomic centers (Figure S3 in SI), thus it is difficult to identify the type of excitation. The rest of the bright states above S3 is dense, producing a broad absorption feature in the short wavelength region. However, if our M06-2X theoretical ECD spectrum is qualitatively correct, it should be experimentally feasible to resolve the S2 and S3 ECD features. So, future experimental work could identify which XC functional, M06-2X or B3LYP,  produce the best absorption and ECD spectra of D-glyceraldehyde.

\subsection{Anisotropy spectra}
\begin{figure}
 \begin{tabular}{cc}
	\resizebox{0.5\textwidth}{!}{\includegraphics{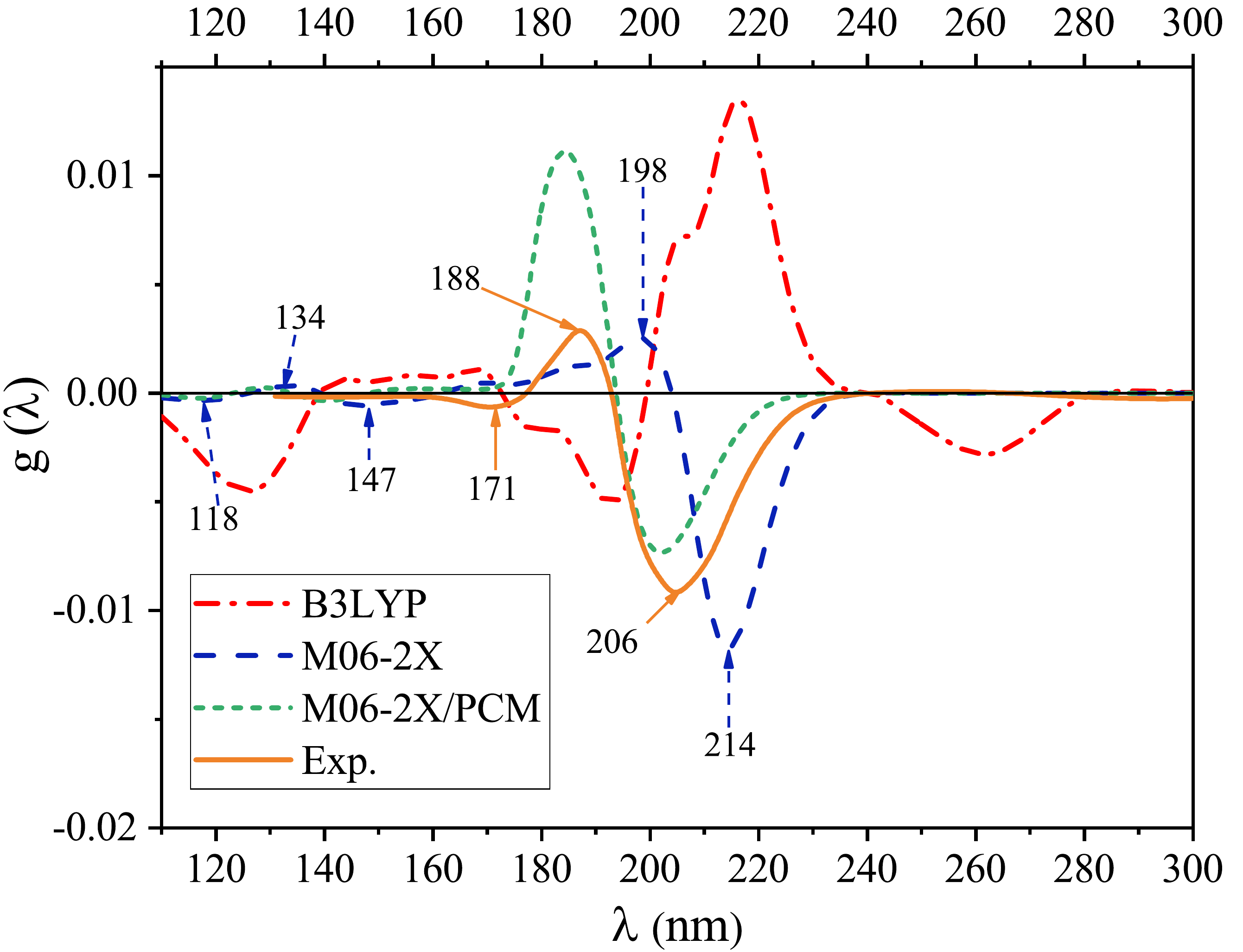}}    &
	\resizebox{0.5\textwidth}{!}{\includegraphics{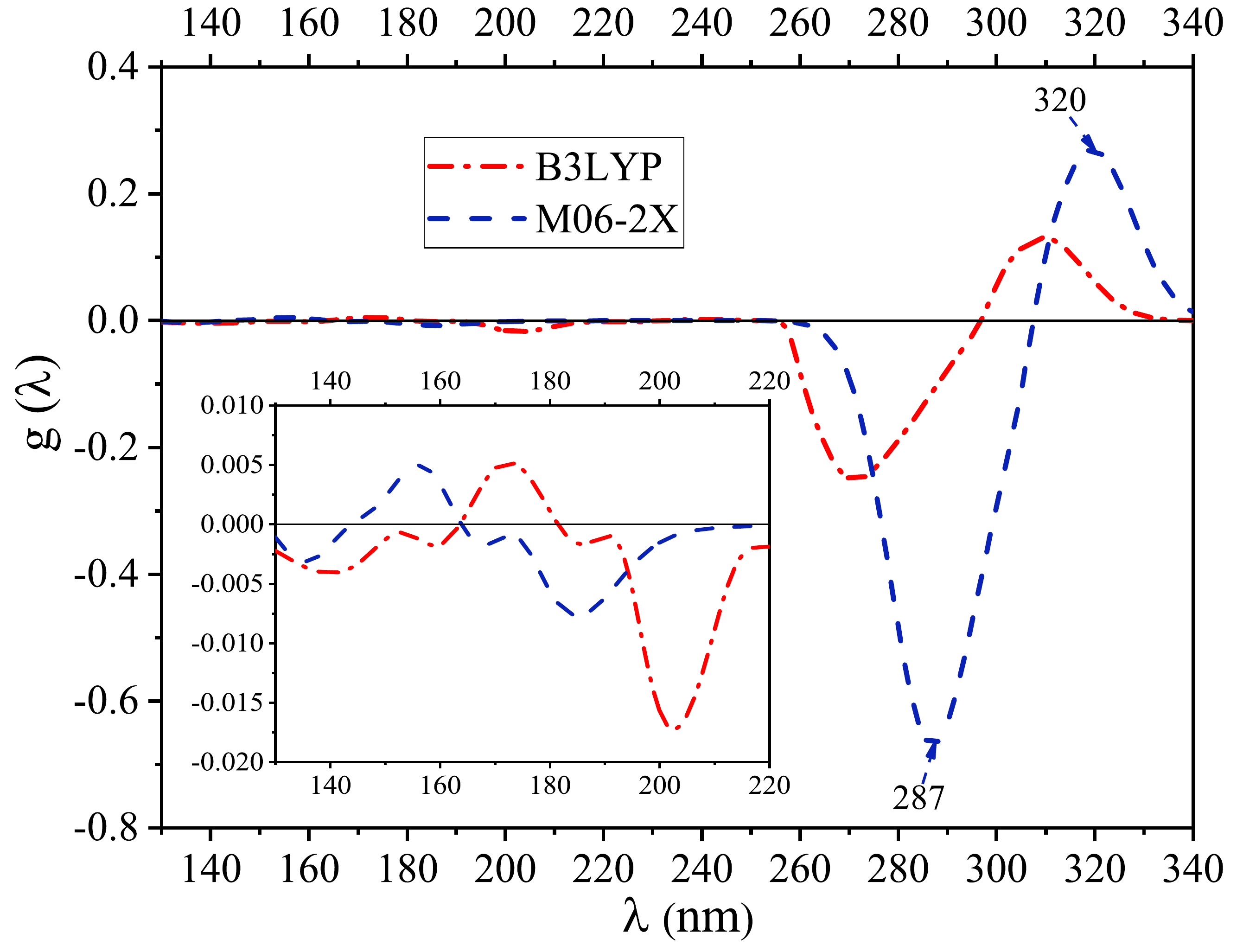}} \\
	(a) & (b) 
     \end{tabular}
	\caption{Anisotropy spectra (a) L-valine zwitterion (b) D-glyceraldehyde. The experimental curve of L-valine  zwitterion is redrawn from Ref. 25. The $g(\lambda)$ values of the theoretical spectra of L-Valine zwitterion are scaled down by a factor of 0.125 for M06-2X and M06-2X/PCM to compare with the experimental result.}
	\label{fig:anispec}
\end{figure}

The wavelength dependent anisotropy factor $g(\lambda)$ of L-valine zwitterion and  D-glyceraldehyde  calculated from TDDFT with the M06-2X and  B3LYP functionals are presented in Figure \ref{fig:anispec}a and \ref{fig:anispec}b, respectively. For L-valine zwitterion, we have also drawn the experimental $g(\lambda)$ and  the anisotropy spectra from TDDFT/M06-2X with PCM calculations in Figure \ref{fig:anispec}a  for comparison.

The anisotropy spectra of the L-valine zwitterion (the blue dash line in Figure \ref{fig:anispec}(a)) calculated without PCM exhibits a high negative peak at $214$ nm and a small positive peak at $198$ nm, which are in good agreement with the experimental peaks\cite{Meinert2012anisotropy} at $206(-)$ nm and $188(+)$ nm, respectively.  Besides, there is a negative peak centered at $171$ nm in the experimental anisotropy spectrum\cite{Meinert2012anisotropy} that should be related to the ECD peak at $165(-)$ nm. Recall, that we have assigned the theoretical ECD peak at $147$ nm as the peak at $165(-)$ nm in experiments,  and thus the negative peak at $147$ nm on the theoretical anisotropy spectra should correspond to the experimental one at $171$ nm. Comparing this small peak with the two peaks in the long-wavelength region, we found that their relative magnitudes are in good agreement with the experimental ones. However, the theoretical spectrum has two peaks situated at $134$ nm and $118$ nm,  whereas we could not identify clearly such extra peaks in the experimental spectrum of  L-Valine  (though, that of D-Valine, which is not the exact mirror image of the former, due to experimental uncertainties, 
displays a small wavy feature in this region)\cite{Meinert2012anisotropy}. This maybe due to the fact that valine films have very strong and broadened absorption in the short-wavelength region,  as indicated by both the theoretical and the experimental study. As a result, it is hard to obtain a highly precise anisotropy spectrum in this region. The spectrum from the TDDFT/M06-2X calculation with PCM also reproduce the first two peaks in the long wavelength region. Although the positions of two peaks agree well with the experimental result, it is not the case of their relative amplitude. The B3LYP functional does not perform well for the valine zwitterion as indicated by the red dash line in Figure \ref{fig:anispec}(a).  In conclusion, the theoretical calculation from TDDFT/M06-2X reproduces the experimental anisotropy spectrum of L-Valine in a satisfactory manner, although the latter was measured in an amorphous film on an MgF2 window.

As seen from Figure \ref{fig:anispec}(b), the anisotropy spectrum of D-glyceraldehyde from the TDDFT/M06-2X calculations is characterized by two high peaks centred at $320$ nm and $287$ nm  with magnitudes of $0.27$ and $-0.70$, respectively. According to the absorption and ECD spectra, the two peaks arise from the first excited state of different conformers. Such large $g$ factors are caused by the much weaker absorption of the S1 state of glyceraldehyde than that of the two lowest excited states of L-Valine. Moreover, a negative peak is located at $184$ nm with a small $g$ factor of $-0.008$. There are many more peaks between $120$ and $180$ nm, however their $g$ factors are small, because the absorption of glyceraldehyde is strong in this region, posing a great challenge for their experimental measurement. The anisotropy spectrum from B3LYP is similar in shape to that of M06-2X, however it is  blue-shifted by  about 10 nm in the short wavelength direction.

\section{Conclusion}
TDDFT calculations with the M06-2X and B3LYP functionals were performed to calculate the absorption, ECD, and anisotropy spectra of the L-valine zwitterion and D-glyceraldehyde. To simulate the experimental spectra, we first carried out conformer searches to hunt for possible low energy structures of the target molecules.  The valine molecule exists in zwitterionic form in the experimentally studied, amorphous film, whereas there is no stable zwitterionic structure in the gas phase. Therefore, the N-H bond length in the \ce{NH3+} group was fixed in the structure optimization and three different geometries were  obtained. Alternatively, the valine zwitterion can be stabilized by modelling the water solvent by means of the PCM. That allowed us to locate another set of three structures that were true local minima of the PES. The analysis of the geometries from the constrained optimization, from PCM and from crystal diffraction, reveals differences of dihedral angles due to variations of inter-molecular interactions. The conformations optimized with constraints were completed with 2 others whose dihedral angles were taken from X-ray structures of valine crystals, and with four more, produced by linear interpolation between the optimized structures and the structures from valine crystals. For D-glyceraldehyde, we have located $32$ structures with no frozen parameter, which were all genuine local minima of the potential energy surface. For both systems, the spectra from all the so-obtained geometries were averaged with Boltzmann weight to calculate their EA and ECD spectra.  

The EA and ECD TDDFT/M06-2X spectra  agree  much better with experimental results than the TDDFT/B3LYP ones. Moreover, this is achieved with no ad-hoc wavelength shifting. We observe also that the TDDFT/M06-2X anisotropy  spectrum of L-valine  reproduces the three main peaks of the experimental spectrum in a satisfactory fashion.  For D-Glyceraldehyde, the excitation wavelengths of the first excited state of $32$ conformers spread from $288$ nm to $322$ nm (TDDFT/M06-2X results), giving rise to two peaks with opposite signs in the ECD spectrum. As a consequence of this and of the very weak absorption of the S1 state, the anisotropy spectra of D-glyceraldehyde exhibits two high peaks at $287$ and $320$ nm, which should be seen in data derived from experiment. Both L-valine and D-glyceraldehyde have large anisotropy factors in the UV region,  a necessary (but, of course, not sufficient condition) for the hypothesis that CP UV-light in space could induce the enantio-enrichment of these life-related chiral molecules.

\begin{acknowledgement}
This work is supported by National Key R{\&}D Program of China (grant No. 2017YFB0203404). B. Suo thank to the financial support from National Natural Science foundation of China (NSFC, grant No. 21673174, 21673175 and 21873077). Dr. Rachel Sparks is acknowledged for exploratory calculations on glyceraldehyde conformers. We thank Dr. C. Meinert for sharing her preliminary experimental results on glyceraldehyde, and the Doeblin F\'ed\'eration for funding.

\end{acknowledgement}

\begin{suppinfo}

\begin{itemize}
 \item The optimized geometries of the valine zwitterion and glyceraldehyde are listed in Tables S1 to S11. Tables S12 to S14 compare the excitation wavelengths and oscillator strength of 30 excited states of the lowest energy geometry of D-glyceraldehyde for different basis sets and different energy functionals. Figure S1 gives the vibrational spectra of glyceraldehyde for its lowest energy geometry optimized by DFT/M06-2X. Figure S2 compares the absorption spectra of glyceraldehyde calculated via TDDFT with five different XC functionals. Figure S3 presents the natural transition orbitals of the S3 state of glyceraldehyde from TDDFT/M06-2X. Figure S4(a) and S4(b) present the ECD spectra of the individual valine zwitterion structures organized according to the sign of the peak with longest wavelength. Figure 4(c) presents the ECD spectra of the individual valine zwitterion structures with PCM for the water solvent.
\end{itemize}

\end{suppinfo}

\bibliography{cdref}

\end{document}